\documentclass[doublecol]{epl2}
\usepackage{graphicx}% Include figure files
\usepackage{dcolumn}% Align table columns on decimal point
\usepackage{bm}% bold math
\usepackage{amsmath,amssymb,amsfonts}
\title{Effect of the orientational relaxation on the collective motion of
patterns formed by self-propelled particles.}
\shorttitle{Collective motion of
patterns formed by self-propelled particles.}
\author{A.I. Chervanyov\inst{1,*} \and H. Gomez\inst{2} \and U. Thiele\inst{1,3}}
\shortauthor{A.I. Chervanyov \etal}
\institute{
\inst{1} Institut f\"ur Theoretische Physik, Westf\"alische Wilhelms-Universit\"at M\"unster,
48149 M\"unster, Germany\\
\inst{2} Departamento de Metodos Matematicos, Universidade da Coruna,
Campus de A Coruna, 15071, A Coruna, Spain\\
\inst{3} Center of Nonlinear Science (CeNoS), Westf\"alische Wilhelms-Universit\"at M\"unster,
48149 M\"unster, Germany\\
\inst{*} Email: chervany@uni-muenster.de
}

\abstract{
We investigate the collective behavior of self-propelled particles (SPPs) undergoing
competitive processes of pattern formation and rotational relaxation of their self-propulsion velocities.
In full accordance
with previous work, we observe transitions between  different steady states of the SPPs
caused by the intricate interplay among the involved effects of pattern formation, orientational order,
and coupling between the SPP density and orientation fields. Based on rigorous analytical and numerical calculations,
we prove that the rate  of the orientational relaxation  of the SPP velocity field is the main factor
determining the steady states of the SPP system. Further, we determine the boundaries between domains in the parameter
plane that delineate qualitatively different resting and moving states. 
In addition, we analytically calculate the collective velocity $\vec{v}$ of the SPPs  and
show that it perfectly agrees with our numerical results. We quantitatively demonstrate that $\vec{v}$
does not vanish upon approaching the transition boundary between the moving pattern
and homogeneous steady states. 
}
\begin{document}
\maketitle
%***********************************************************************
\section{Introduction}
One of the most intriguing aspects of the collective behavior of a system of  self-propelled (active)
particles (SPP) \cite{hydroactive}, is that the  motion of each individual particle can cause
an organized ("collective") motion of the SPP system as a whole. At high concentrations of SPPs,
the described collective motion is known to occur in systems of active colloids that form regular patterns
termed "active crystals" in \cite{LowenPRL1}. The presence of active drive makes the properties of the
described "active crystals" very different from  crystals formed by ordinary "passive" colloids.
In particular, self-propulsion of colloidal particles can significantly lower \cite{LowenPRL2,LowenPRL3}
the concentration threshold of the entropy-driven  clustering relative to that observed in 
the system of passive colloids. 
Moreover, an additional energy associated
with the  individual motion of the SPPs can be consumed not only by shifting the crystallization point of
the SPP system, but also by inducing the organized motion of the SPPs. In this scenario, the orientational interactions between
the individual SPPs can cause an essentially non-equilibrium state, where the SPP system moves in a direction chosen by symmetries
of the patterns formed by these SPPs. The described states of collective motion have been observed experimentally in many
relevant biological
systems, ranging from  bacteria colonies \cite{micro,micro1,micro2,micro3} to the flocks of 
birds \cite{birds, TonerTu}, which explains the high interest in
understanding the physical effects underlying these states. Despite
essential theoretical progress \cite{hydroactive} in the field achieved over the last decade, many aspects of the described collective
behavior of SPPs still remain rather obscure. One of such key aspects
addressed in the present work, is the relative significance of the tendency of SPPs to form patterns and of the orientational relaxation
of their velocities for the formation of moving and resting steady states of the SPP system.

By making use of the minimal model described below, we  demonstrate that the orientational relaxation of the
SPP velocities is the main factor playing in favor of suppressing the collective motion of the SPPs caused by their self-propulsion.
More specifically, we will show that the transition between
the observed moving ordered state, resting ordered state and disordered (homogeneous) state can be induced solely by changing the rate of the orientational
relaxation of the SPPs. According to our exact results described below, the individual motion of SPPs  {\it always}
causes a tendency towards collective motion of the whole SPP system.  In this sense, there exists no single threshold magnitude
of the active drive that delineates the regimes of moving and resting states. The main factor that suppresses the collective motion of the SPPs
is a sufficiently quick orientational relaxation of the local orientation of the SPP velocities quantified by the "polarization" field 
$\mathbf{P}(\vec{r})$.

\section{The model}
In order to rationalize the described interplay between the tendencies of the active SPP system to the crystallization and collective motion,
it is instructive to study a modification of the well-established colloid crystallization model that would adequately account for the active
drive of the colloids. A minimal model of that type has been recently developed in \cite{LowenPRL1}, by combining the Phase-Field-Crystal
(PFC) model \cite{ElderGrantPRE,UwePRE,EmmerichLowen} and Toner-Tu (TT) \cite{TonerTu} approach describing the kinetic flocculation with 
the effect of the
orientational order of flocks taken into account. Owing to its transparency and simplicity, this model provides an excellent framework
for studying the complicated interplay between the spatial and orientational ordering of the SPPs. Although this pioneering work provides very
useful insight  into the nature of the collective behavior of SSPs, it does not give a systematic account for possible transitions between
moving and resting  states of SPPs depending on the relative significance of the involved effects. In the present work, we 
use a slightly modified version of the described model with the objective to construct the exact "phase diagram" describing all the possible
collective states of the SPPs,
as well as the transition boundaries between them. In addition, we  investigate, both analytically and numerically, the dependence of the
velocity of the collective motion on the involved parameters. As we demonstrate below, this detailed investigation leads to
a new understanding of the relative role of the involved effects (e.g. pattern formation and rotational diffusion) in the formation of the
observed moving and resting patterns in the SPP system.

Having implemented the above outlined strategy of combining the PFC and TT approaches,  we employ
the above outlined idea of describing the correlations between the density $\rho(\vec{r})$ and orientational $\vec{P}({\vec{r}})$ fields 
with help of coupled PFC and TT equations. Note that the main quantity of interest $\rho$ termed "density" in the above
for the sake of convenience,
is assumed to have no limitations with respect to the domain of values it can assume. Similarly as in the standard PFC approach, this quantity
should be understood as an order parameter (e.g. a density excess over the critical density).  Therefore, $\rho$ can assume, 
in particular, negative values. Note that  additional limitations
on the domain of possible values of $\rho$ imposed by specific physical conditions can be easily incorporated into
the described model without causing a loss of its generality. One such possibility arises from properly defining the average density
$\bar{\rho}$, considered an input parameter of the model.

For the sake of tractability, we use a simplest possible form of the  dynamical equation for the  vector quantity   $\vec{P}$,
termed "polarization" hereafter,
describing the orientational order in the field of the SPP velocities. This equation of the form
\begin{equation}
 \label{1}
\frac{\partial\vec{P}}{\partial t} =  -\tau^{-1}\left(1-\gamma\nabla^2\right)\vec{P} - v_0\vec{\nabla}\rho
\end{equation}
has been first empirically obtained in \cite{Fily}, by making a comparison with the results of the simulations for  SPP disks.
Note that the terms on the right hand side (r.h.s.) of Eq.(\ref{1}) containing $\vec{P}$ can be obtained by dropping 
the non-linear terms in the original TT equations. The generalization of Eq.(\ref{1}) containing additional non-linear terms 
has been independently obtained in \cite{LowenPRL1}.
Coupling of the orientational order to the
density field is described by the last pressure-like gradient term in the r.h.s. of Eq.(\ref{1}) featuring the constant
magnitude $v_0$ of the self-propulsion velocity of the SPPs. The coefficients $\tau^{-1}$ and $\gamma$  quantify,
respectively, the relaxation rate due to the orientational diffusion and significance of the spatial dispersion of the polarization.

Note that the simple diffusion equation for density $\rho$ used in \cite{Fily} is not suited for describing 
the formation of periodic patterns of the SPP density needed to mimic the crystallization of SPPs in our approach. 
In order to improve this feature of the model, critical for its applicability
to the case of active crystals under consideration, we follow  \cite{LowenPRL1} and employ the PFC 
equation for the density $\rho(\vec{r})$ of the form
\begin{equation}
 \label{2}
\frac{\partial\rho}{\partial t} = \nabla^2 \left[\omega(\nabla)\rho + (\bar{\rho}+\rho)^3  \right] -
v_0 \vec{\nabla}\cdot\vec{P},
\end{equation}
where we have separated out the average density $\bar{\rho}$ for the sake of convenience. Here,
\begin{equation}
 \label{2.1}
 \omega(\nabla) = -\epsilon + \left(q^2+\nabla^2\right)^2
\end{equation}
is the $4$th-order differential operator with the only control  parameter $\epsilon$ (reduced deviation from the critical temperature), inherited
from the standard PFC model \cite{ElderGrantPRE,EmmerichLowen,Steinbach}. Recall that in this model the positive (negative) values of $\epsilon$ 
correspond to the
ordered patterned (homogeneous) steady states of the density field. The morphology of patterns formed at positive $\epsilon$ is determined by 
the average
density $\bar{\rho}$. Note that the only difference that distinguishes the standard PFC equation from its modification in Eq.(\ref{2}) is the
last term in its r.h.s. describing the coupling of the density structure to the polarization field.

It is important to note an essential difference between the structures of the equations in Eq.(\ref{2})
and Eq.(\ref{1}). Eq.(\ref{2}) is a non-linear equation for the density coupled to  Eq.(\ref{1}) through
the linear coupling term $v_0 \vec{\nabla}\cdot\vec{P}$. Eq.(\ref{1}) is, in contrast, a linear differential equation,
which greatly simplifies the analysis of the coupled dynamics of the fields $\rho$ and $\vec{P}$. Specifically,
in the frameworks of the proposed linear model for the dynamics of $\vec{P}$, the effect of the orientational field on
the pattern formation and collective motion of SPPs can be fully understood in terms of the relation among the
relaxation rate  $\tau^{-1}$  and reduced self-propulsion velocity $v_0$. In the extreme case of quick relaxation of the
orientational field $\vec{P}$ ($q\tau v_0\ll 1$), the spatial structure of the polarization field $\vec{P}$ immediately adjusts to the
structure of the density field $\rho$ of the SPPs. Physically, this regime corresponds to sufficiently large undercooling, where 
the local mobility of the SPP domains depends only on their density, so that the most "crowded" domains have the smallest local velocities. 
In the symmetric density phases, the individual motion of the SPP domains moving in the opposite directions averages out,
thus resulting in the resting state. In this case, the orientational relaxation of $\vec{P}$ 
causes no qualitative difference in the spatial structure of $\rho$ relative to the predictions of the standard PFC
model. As will be shown in what follows, the described quick relaxation of $\vec{P}$  causes only a quantitative
correction to the PFC control parameter $\epsilon$ in Eq.(\ref{2.1}) that leads to a shift of the phase boundary separating the domains
of patterns and homogeneous SPP phase. In the opposite case $q\tau v_0\gg 1$ of slow orientational relaxation, a more complicated 
response
of the density field to the orientational relaxation is observed. In particular, slow relaxation of $\vec{P}$ results in the occurrence of 
steady states
in the form of moving patterns. In contrast to the former case of the quick
relaxation of $\vec{P}$, this case cannot be reduced to any modification of the PFC model.

\begin{figure}
\centering
\includegraphics[width=8.5cm,angle=0]{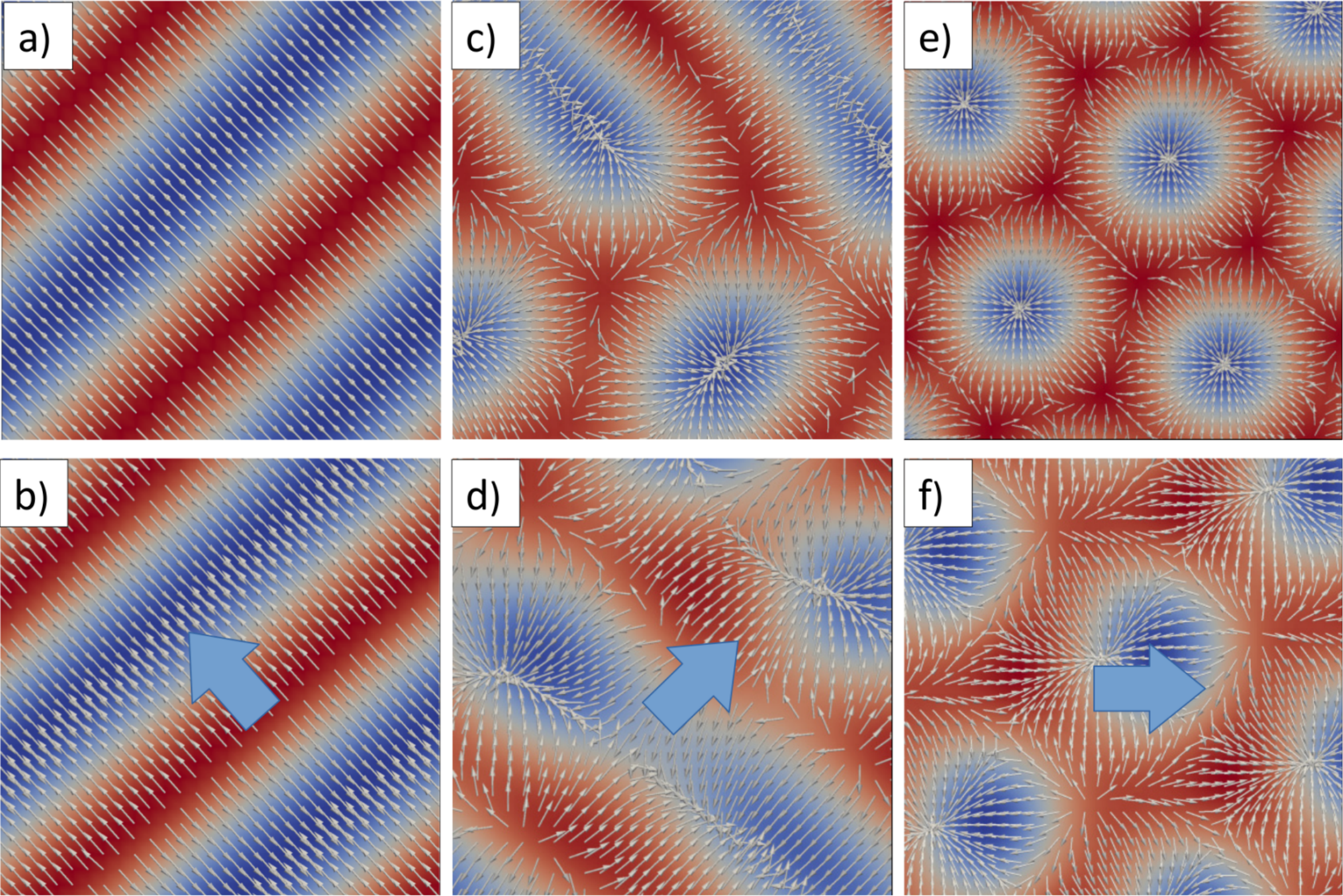}
\caption{\label{fig1} Examples of the morphologies of resting (a, c, e) and moving (b, d, f) patterns obtained from the performed 
iso-geometric finite element analysis of Eqs.(\ref{1}-\ref{2}) for the selected values  of the control parameters. 
In all the examples
$\epsilon = 1$, $v_0 = 0.3$, $\gamma = 0$. Left column: $\bar{\rho}=0$, a) $\tau^{-1} = 0.30$, b) $\tau^{-1} = 0.20$.
Middle column: $\bar{\rho}=0.1$, c) $\tau^{-1} = 0.30$, d) $\tau^{-1} = 0.20$. Right column: $\bar{\rho}=0.2$, 
e) $\tau^{-1} = 0.30$, f) $\tau^{-1} = 0.20$. The blue arrows show the directions of the collective drift of the SPPs.
The glyph vectors indicate the direction and magnitude of the polarization field $\vec{P}(\vec{r})$.}
\end{figure}

\section{Linear stability analysis}
The above outlined qualitative picture can be rationalized by employing an analytical linear stability analysis of the simultaneous 
Eqs.(\ref{1}-\ref{2}). The obtained analytical prediction for the phase-diagram of pattern formation
in the system of SPPs are found to be in perfect agreement  with that obtained from the numerical solution of the same equations. We attribute this good
agreement to the linear
character of Eq.(\ref{1}) that captures the effects of the orientational field on the density structure on the linear response level.

Linear stability analysis needs to be performed by taking into account the actual morphology of the patterns occurring in the SPP system
under investigation. The possible morphologies are determined (see Fig.1)  by numerically  solving 
Eqs.(\ref{1}-\ref{2}) as described below. These morphologies are found out to be largely the same as those observed in the standard PFC model. 
For the sake of brevity, we
do not discuss different morphologies (e.g. checkerboard and zig-zag lamellae morphologies) that have been observed by us only in the extreme 
cases of
strong coupling $v_0\sim 1$ or for large values of  $\epsilon \sim 3$. Neither do we analyze non-periodic patterns that 
occur as "metastable" states between the lamellae and hexagonal morphologies (see Fig. 1 c, d). It turns out that
the morphology of the patterns  affects neither  the collective velocity nor the shift in the crystallization temperature. This 
observation  gives a rational for restricting our analysis to the cases of lamellae and hexagonal morphologies known from 
the standard PFC model.

In addition to the well studied resting steady states occurring in the standard PFC, one needs to take into account the possibility 
that the patterns can move with constant velocity $v$, which complicates the associated mathematical development.
The simplest observed one-dimensional morphology, the lamellae morphology (see Fig.~1~a,b)),  is comprised of  parallel stripes (rolls) that
show no spatial variation along their axes.
The structure of the lamellae morphology of a given periodicity $\frac{2\pi}{|\vec{k}|}$ can therefore be described by a single wave vector
$\vec{k}$ perpendicular to the stripes.  For the sake of simplicity, we chose to work in the coordinate frame associated with
the wave vector $\vec{k}$ pointing in the direction of the $x$-axis of this frame. With the described choice of the coordinate frame, 
a set of equations given by Eqs.(\ref{1}-\ref{2}) reduces to two coupled equations for $\rho$ and $P_x$ with  $P_y = 0$. Periodic modes 
of the density and orientational 
fields corresponding to the lamellae morphology moving with velocity $\vec{v}$ can therefore be described by a linear 
combination of the plane waves $\sim \exp{(\pm ik(x\pm |\vec{v}|t))}$ having different 
amplitudes and wave numbers $k =|\vec{k}|$. 
An important feature of the  fields
$\rho$ and $P_x$ that needs to be taken into account in the performed linear analysis, is the phase shift between these 
periodic fields that can be elucidated by a comparison of the numerical results shown in Figs. 1a) and 1b). We describe this shift by 
employing the following
forms of the fields $\rho\sim \cos{(k(x-vt))}$ and $P_x \sim \sin{(k(x+\Delta x-vt))}$. Substituting these expressions into
the simultaneous equations Eqs.(\ref{1}-\ref{2}), linearizing with respect to the amplitudes of  the fields $\rho$ and $P_x$,
and deriving the consistency conditions for the obtained algebraic linear homogeneous equations for the amplitudes, one arrives at

\begin{align}
 \label{15}
 (v^2-v_0^2 -v(k\tilde{\tau})^{-1} \tan{(k(\tilde{x}+\Delta x))})\tan{(k\tilde{x})}= \nonumber \\ 
(\tilde{\epsilon}-(k^2-q^2)^2)(vk-\tilde{\tau}^{-1}\tan{(k(\tilde{x}+\Delta x))}) ,
\end{align}
where $\tilde{\epsilon}=\epsilon - 3\bar{\rho}^2$, $\tilde{\tau}=\tau (1+\gamma k^2)^{-1}$
and $\tilde{x} = x - vt$. Note that due to the 
presence of the gradient term in the operator 
$\omega(\nabla)$ in Eq.(\ref{2}), the described linearization of Eqs.(\ref{1}-\ref{2}) always admits the additional 
constant solution corresponding to the wave vector $k=0$. Hereafter, we omit this trivial solution that is inherent to
the  nature of the used "conserved" PFC equation that preserves the average SPP density.

Eq.(\ref{15}) admits the solution $\Delta x = 0$, $v = 0$,
\begin{equation}
 \label{4}
 \tilde{\epsilon} = v_0^2\tilde{\tau} + (k^2-q^2)^2,
\end{equation}
which describes the stationary (resting) steady state. It is important to note that
this case corresponds to the zero phase shift $\Delta x  = 0$. This observation quantitatively corroborates 
the above qualitative consideration that the field of the polarization is in phase with the gradient of the density field
in the stationary steady state of the SPP system. 
In this case, the solution to the linearized equations obtained from Eqs.(\ref{1}-\ref{2})  is fully determined by the dispersion 
relation given by Eq.(\ref{4}).

In the case of non-zero velocities of the collective motion, Eq.(\ref{15}) reduces to the 
explicit expressions for the dispersion relation $\tilde{\epsilon}(k)$, collective velocity $v$ and phase shift $q\Delta x$ 
of the form
\begin{eqnarray}
 \label{5}
 \tilde{\epsilon} = k^{-2}\tilde{\tau}^{-1} + (k^2-q^2)^2, \quad v^2 = v_0^2 - k^{-2}\tilde{\tau}^{-2}, \\
 \tan{(k\Delta x)} =kv\tilde{\tau}. 
 \end{eqnarray}

The dispersion relations $\tilde{\epsilon}$ given by Eq.~(\ref{4}) and Eq.~(\ref{5}) determine the non-trivial values of the critical 
wave numbers that 
delineate the regions of stability and instability of the solutions of Eqs.~(\ref{1}-\ref{2}), corresponding to 
the resting and moving steady states, respectively. According to these relations, the obtained solution of the linearized dynamical 
equations 
is unstable in a  band 
of wave numbers in the vicinity of $k=q$. In the presence of  non-linear term $\sim (\rho + \bar{\rho})^3$ 
in the r.h.s. of Eq.(\ref{2}), 
the exponentially growing modes, causing the instabilities, 
become suppressed so that they saturate to a finite 
limit with increasing the time. Similarly as in the standard PFC \cite{UwePRE}, this suppression results in shrinking 
the described instability band to a single
critical wave number $q$ that corresponds to the only stable mode. Near the onset of the non-linear instability, one can
therefore expand the r.h.s. of Eq.(\ref{4}) and Eq. (\ref{5}) at $k=q$. Up to the leading order these expansions read,
\begin{equation}
 \label{6}
\tilde{\epsilon}_R = v_0^2\tau (1+\gamma q^2)^{-1}, \quad
\tilde{\epsilon}_L = q^{-2} \tau ^{-1}(1+\gamma q^2) 
 \end{equation}
 \begin{equation}
 \label{10}
 v^2 = v_0^2 - q^{-2}\tau^{-2}(1 + q^2\gamma )^2,
 \end{equation}
respectively.

Note that the above calculation is valid only for the simplest case of the one-dimensional roll patterns.  
In order to obtain the dispersion relation
and collective velocity for more complicated two-dimensional pattern morphologies, one needs to use a linear combination of the 
considered roll solutions,
corresponding to symmetries of these morphologies. For the case of the hexagonal patterns shown in Fig. 1 e),f),
an appropriate  linear combination of the roll solutions, admitted by the linearization of Eqs.(\ref{1}-\ref{2}),  reads
\begin{equation}
\label{6.1}
 \sum\limits_{l=1}^{3} r_l\cos{(k \vec{k}_l\cdot \vec{\tilde{r}})},\quad
 \sum\limits_{l=1}^{3} p_l^{x,y}\sin{(k \vec{k}_l\cdot (\vec{\tilde{r}} + \Delta\vec{r}_l))},
\end{equation}
 where $r_l$, $p_l^j$ are, respectively, the amplitudes of the density field and the $j$th component of the  polarization  field,
$\vec{\tilde{r}} = \vec{r} - \vec{v}_l t$, and $k\Delta\vec{r}_l$ are the phase shifts of the polarization field.
 The vectors $\vec{k}_l$, associated with
 different roll modes that form hexagons upon interference, are given by $\vec{k}_1= \vec{e}_x$, $\vec{k}_2= (-\vec{e}_x +\sqrt{3}\vec{e}_y)/2$,
 and $\vec{k}_3= (-\vec{e}_x - \sqrt{3}\vec{e}_y)/2$, $\vec{e}_x$ and $\vec{e}_y$ being the position vectors associated with a chosen Cartesian
 coordinate frame. Note that $|\vec{k}_1|=|\vec{k}_2|=|\vec{k}_3|=1$. Applying the procedure used above for the case of a single roll solution
 to each of the exponential
 terms given in the r.h.s. of Eq.(\ref{6.1}), one arrives at a set of three consistency conditions that lead to an explicit expression for the
 dispersion relation, velocities $\vec{v}_l$ and phase shifts $\Delta\vec{r}_l$. The thus obtained dispersion relations $\tilde{\epsilon}(k)$ 
 prove to be the same
 as those given by Eq.(\ref{6}) for the one-roll solution in the cases of zero ($\epsilon_R$) and non-zero ($\epsilon_L$)
 collective velocities, respectively.
 The velocities of the rolls are obtained from the above consistency conditions in the form $\vec{v}_l=  v \vec{k}_l$,
 with  $v$ given by Eq.(\ref{10}). Eq.(\ref{6.1}) therefore describes  the three rolls, independently moving with the same velocity in the 
 directions
 normal to their axes. Note that for the considered hexagonal morphology, the quantity $v$ describes the reduced
 temporal frequency $qv$ of changing  the hexagonal patterns in time, and has been called "velocity" only for the sake of unifying
 the description with the one-dimensional case. The phase shifts are obtained in the form $\Delta\vec{r}_l = \Delta x \vec{k}_l$ with $\Delta x$
 given by Eq.~(7). 
\begin{figure}
\centering
\includegraphics[height=5cm,angle=0]{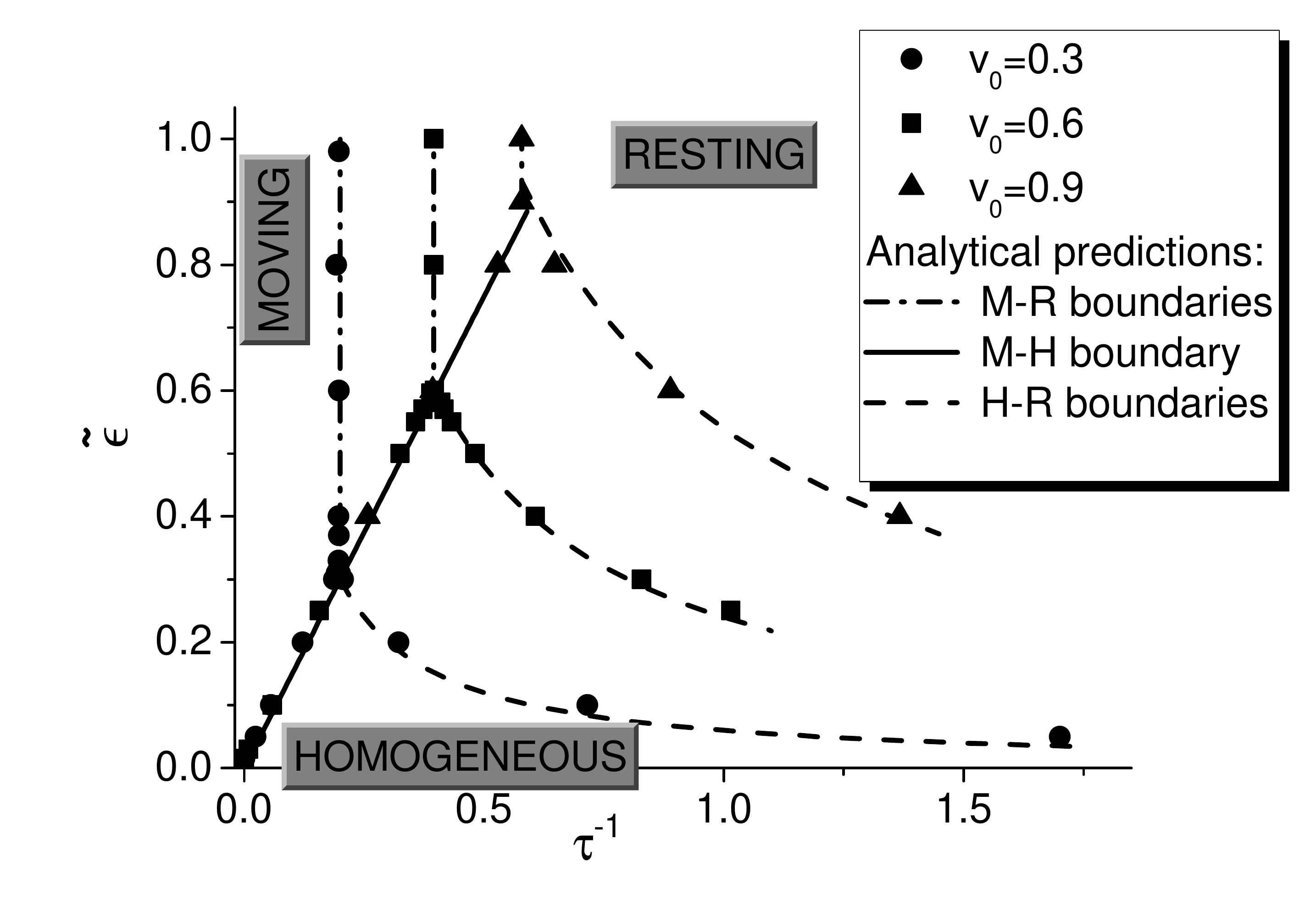}
\caption{\label{fig2} Quasi-phase diagram of moving and resting SPP steady states for the selected values of the coupling  $v_0$ 
in terms of the rate $\tau^{-1}$ of the orientational relaxation and the reduced control parameter $\tilde{\epsilon}$ for 
$\gamma = 0.5$, $q=1$: 
comparison of the analytical predictions by Eqs.(\ref{5}-\ref{6}) (solid line) and  numerical results (scatter points).
}
\end{figure}

Two important observations regarding the obtained analytical results given by Eqs.(\ref{5},\ref{6},\ref{10}) are in order here. 
First,  the r.h.s. of the
expressions for the critical control parameter given by  Eq.(\ref{6}) does not depend on the
average density $\bar{\rho}$. Recall that parameter $\bar{\rho}$ in the PFC equation for the SPP density given by Eq.(\ref{2}),
is responsible for controlling the morphology of
the patterns. Therefore, the above observation speaks in favor of the statement, alluded in the introductory part, that 
the shape of the boundaries delineating the steady states of the SPP system does not depend on the morphology of the SPP patterns. 
The second important observation regarding Eq.(\ref{10}) is that the velocity $v$ of the pattern motion  does not depend on the control 
parameter $\epsilon$. It is interesting
that according to the results of the numerical calculations, this property of $v$ proves to hold not only near the onset of 
instability, but in the whole explored domain of values of the involved control parameters.

The derived expressions for the reduced control parameter $\tilde{\epsilon}$ and the velocity $v$ near the onset of
instability are the central results of the present work. In order to verify these analytical findings, we perform
isogeometric finite element analysis \cite{Hector1,Hector2} of Eqs.(\ref{1}-\ref{2}) over a wide
range of the parameters $\epsilon$, $v_0$,  $\tau$ and $\gamma$. In order to eliminate finite-size effects
we impose the standard \cite{Hector1} periodic boundary conditions and use a relatively large ($72.7 q^{-1}\times 72.7 q^{-1}$) simulation domain size.  
We have "fixed"  the morphology
of the SPP patterns by giving appropriate values to the average density $\bar{\rho}$. Specifically, we have used the three
values $\bar{\rho} = 0.0; 0.2; 0.1$ that correspond to the lamellae (roll), hexagonal and "metastable" lamellae-hexagonal patterns,
respectively. The described patterns, obtained from the simulations, are shown in Fig.1. Note that the obtained patterns are found 
to move or rest at the respective values of the orientational relaxation rate  $\tau^{-1}$ predicted analytically.
By fixing chosen values of the parameters $\bar{\rho}$, $v_0$ and $\gamma$ and varying the values of $\tau$ and $\epsilon$, we determine
morphologies  and velocities (if any) of the corresponding patterns formed by SPPs. As a result, we identify the phase boundaries separating
different regimes (e.g. moving and resting patterns) in the $\tau^{-1}-\epsilon$ plane. The results of this analysis are depicted in Fig.2-3.
As can be seen from these Figures, our simple analytical expressions given by Eqs.(\ref{5}-\ref{10}) provide a perfectly adequate prediction 
for the boundaries between the possible steady states and the collective velocity of the SPP system. The determined steady states of the SPP system 
and the transition boundaries between them are represented in the form of a "phase diagram" described in the next section. 

\section{Resting and moving states of the SPP system}
Based on our analytical and numerical results, we have identified three different regimes of the qualitatively different behavior of SPPs: 
homogeneous state,
moving patterns and resting patterns. Note that this result is in qualitative agreement with \cite{LowenPRL1}. Our more detailed analysis leads, 
however,
to several essentially novel features relative to this previous work, critical for understanding the behavior of the SPPs. First of all, we emphasize 
that the moving state of the patterns occurs for any non-zero magnitude of the self-propulsion velocity $v_0$ at sufficiently small values of the
orientational relaxation rate $\tau^{-1}$.
According to the performed analysis,
increasing the relaxation rate $\tau^{-1}$ at any given $\epsilon>3\bar{\rho}^2$ causes the transition from the "moving" to the "resting"  state 
of the patterns.
The nature of this transition depends on the value of the control parameter $\epsilon$. Above certain critical value $\epsilon_c$, the described
moving-resting pattern ($M-R$) transition occurs directly, without crossing over the domain of the homogeneous SPP state. For $\epsilon<\epsilon_c$,
in contrast, increasing $\tau^{-1}$ causes a sequence of two transitions: (i) moving pattern-homogeneous state ($M-H$) transition  and 
homogeneous state- resting pattern ($H-R$)
transition.
The critical value $\epsilon_c$ can be easily evaluated by determining the intersection of the dispersion curves $\epsilon_L(\tau^{-1})$ and
$\epsilon_R(\tau^{-1})$ described by the corresponding dispersion relations given by Eq.(\ref{6}). The result of this evaluation leads to an
incredible simple result $\tilde{\epsilon}_c = v_0 q^{-1}$, which gives excellent agreement with the numerical calculation (see Fig.2). In the
$\tau^{-1}-\epsilon$ diagram shown in Fig.2, the domains corresponding to the homogeneous state ($H$), resting ($R$) and moving ($M$) patterns, 
are separated by the three different
phase boundaries $M-R$, $M-H$, and $H-R$. Analytical predictions for these  boundaries  
are represented by the dash-dotted, solid and dashed curves respectively defined by the equations $\tau^{-1} = \frac{qv_0}{1+q^2\gamma}$
($\tilde{\epsilon}\ge v_0 q^{-1})$,  $\epsilon= 3\bar{\rho}^2 + \tilde{\epsilon}_L$, and $\epsilon= 3\bar{\rho}^2 + \tilde{\epsilon}_R$ with 
$\tilde{\epsilon}_{R,L}$
given by Eq.~(\ref{6}). The scatter points in the diagram show the numerical results for the described phase boundaries. 
In order to ensure a high quality of the verification of the obtained analytical expressions, the phase boundaries have been
determined numerically with the accuracy $\Delta\tau^{-1} = \pm 10^{-3}$.
As can be seen from Fig. 2, the described numerical calculations and analytical predictions for the phase diagram  show excellent
agreement. This allow us to conclude that the linear stability analysis of the coupled non-linear PFC-TT equations provides quite a 
sufficient description of the transitions between the steady states of the SPP system.

According to the obtained phase diagram shown in Fig.2, in the case when the effect of the orientational relaxation is not sufficient to prevent
the SPPs from collective motion, the patterns move with the constant velocity $v$. It is instructive to investigate the dependence of
this velocity of the patterns on the involved parameters $\epsilon$, $\bar{\rho}$, $\tau$ and $\gamma$. Illustrative results of the
numerical calculations of $v$ in different domains of the control parameter $\epsilon$ are shown in Fig.3. As can be seen from this Figure,
the obtained numerical predictions are in a good agreement with the analytical expression given by Eq.(\ref{10}). Moreover, this expression 
 proves to perfectly describe the numerically obtained dependence $v(\tau^{-1})$ 
in the whole investigated range of $\tau^{-1}$. Note that $v(k)$ in Eq.(\ref{5}), obtained from the linearized dynamical equations, 
does not depend on the parameters $\epsilon$ and $\bar{\rho}$. The only approximation that is used to derive
$v(q)$ in Eq.(\ref{10}) from its "linear" counterpart $v(k)$ in Eq.(\ref{5}) consists in selecting a specific value of the wave number $k=q$ 
that corresponds to the only stable solution of the nonlinear equations. This approximation does not specifically rely on the expansion 
near the transition boundary (i.e. at $\epsilon \sim \tilde{\epsilon}_L+3\bar{\rho}^2$), which explains the observed validity of 
Eq.(\ref{10}) away from this boundary. In addition, $v$ given by Eq.(\ref{10}) does not depend on the parameter
$\bar{\rho}$, so that the magnitude of the velocity is not affected by the morphology of the patterns.

Further, for any $\epsilon > 3\bar{\rho}^2$ there exists a finite range
$0\le\tau^{-1}\le (\epsilon - 3\bar{\rho}^2)(\gamma + q^{-2})^{-1}$ where the SPPs experience a collective motion. The velocity of this motion
monotonically decreases with decreasing $\tau$ from the maximum value $v_0$ to the minimal value reached at the transition boundary.
An important feature of the obtained dependance $v (\tau)$ shown in Fig.3 is that it behaves in qualitatively different ways upon
approaching the phase boundaries corresponding to the $M-H$ and $M-R$ transitions.
In the domain of small control parameters $\tilde{\epsilon} < q^{-1}v_0$ corresponding to the $M-H$ transition,
$v$ monotonically decays to a finite
value that evaluates to $\sqrt{v_0^2-q^{2}\tilde{\epsilon}^2}$, as predicted by Eqs.~(\ref{6})-(\ref{10}) (see Fig 3a). 
In the domain $\tilde{\epsilon} > q^{-1}v_0$, the velocity 
tends to zero upon
approaching the $M-R$ phase boundary (see Fig 3b). In order to correlate the observed behavior of the pattern velocity $v$ with the structure of the
patterns, we have numerically calculated the root-mean-square-deviation (RMSD) defined as
$r_m (t,\tau) = (V^{-1}\int (\rho(\vec{r},t)-\bar{\rho})^2 d^2 r)^{1/2}$, $V$ being the area of the computational domain.  RMSD
describes the spatial dispersion of the SPPs at a given moment of time $t$. As can be seen from the results of the calculation of
the asymptotical values of $r_m (t)$ at  $t\to\infty$ shown by the squares in Fig.3, $r_m (\infty,\tau)$ shows just the opposite trend
relative to $v(\tau)$. Specifically, upon approaching the phase boundary, $r_m (\infty,\tau)$ tends to a finite value for 
$\tilde{\epsilon} > q^{-1}v_0$, and it decays to zero for $\tilde{\epsilon} < q^{-1}v_0$.
\begin{figure*}
\centering
\includegraphics[height=5cm,angle=0]{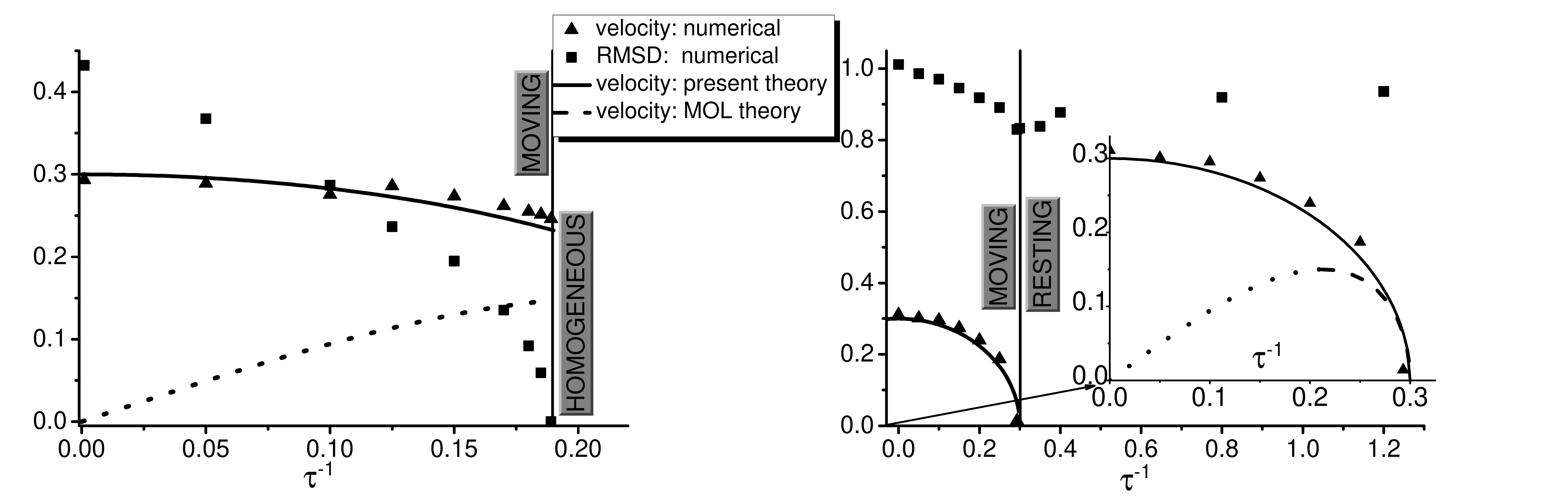}
\caption{\label{fig3} Velocity of the collective motion $v$ (triangles) and root-mean-square deviation RMSD (squares)  of the 
density as functions of the orientation relaxation rate $\tau^{-1}$ for $\bar{\rho} = 0$, $v_0 =0.3$, $q=1$ and a)  $\epsilon = 0.2$;
b) $\epsilon = 1.0$: comparison of our analytical prediction given by Eq.(\ref{10}), analytical 
prediction of the MOL theory in \cite{LowenPRE1} close to the M-R transition (dashed) and away from its formal applicability 
region (dotted), and the results of the numerical analysis. Inset of b) shows a zoom
depicting the behavior of $v(\tau^{-1})$ at the $M-R$ boundary.}
\end{figure*}

As can be elucidated from Fig.3 and our result in Eq.(\ref{10}), in the absence
of the rotational diffusion and dispersion ($\tau^{-1}=\gamma = 0$), the patterns move with the constant velocity $v_0$.
This collective motion is inherent to the individual motion of the SPPs with the same self-propulsion velocity $v_0$,
not affected by the above described "synchronization" of the local density and orientation  fields.
The occurrence of the rotational diffusion and dispersion ($\tau^{-1}\ne 0, \gamma\ne 0$)  plays in favor of  suppressing
this motion.  This suppression results in the  monotonic decay
of the velocity $v$ with increasing the rate of the rotational diffusion predicted by our result given by Eq.(\ref{10}).

The above described monotonic behavior of the function $v(\tau^{-1})$, which proves to be the same for the one-dimensional
roll patterns and two-dimensional hexagonal patterns, shows qualitative differences with its counterpart $v_m$ obtained
in  \cite{LowenPRE1} for the one-dimensional case. After adopting  the notations used in \cite{LowenPRE1} to
those employed in the present work, $v_m$ can be written as
\begin{equation}
 \label{11}
 v_m^2 = q^{-2}\tau^{-2}(1+q^2\gamma)^2 - v_0^{-2}q^{-4}\tau^{-4}(1+q^2\gamma)^4.
\end{equation}
The prediction for the collective velocity $v_m$ given by Eq.(\ref{11}) is depicted by the dashed curve and is referred to as 
"MOL theory" in Fig.3. Note that according to \cite{LowenPRE1}, the validity of  Eq.(\ref{11}) "remains restricted 
to the very close vicinity of the threshold for the onset of collective motion". Our numerical results, depicted in 
Fig.~3 by triangles, corroborate this statement only for the case of the $M-R$ boundary (i.e. at $\tilde{\epsilon}>q^{-1}v_0$)
shown in Fig.~3b).  Comparison of the prediction of Eq.(\ref{11}) against our numerical results 
makes it possible to establish quantitative criteria for the qualitative validity of MOL theory, to be expressed as
$\tau^{-1}>\tau_{max}^{-1}\equiv 2^{-\frac{1}{2}}q(1+q^2\gamma^2)^{-1}v_0$; $\tilde{\epsilon}>q^{-1}v_0$.
Outside of this range of its validity, an overdrawn application of  Eq.(\ref{11}) would lead to predicting an increase  
in the collective velocity $v(\tau^{-1})$ as the rate of the rotational diffusion $\tau^{-1}$ increases, as well as vanishing 
$v$  with vanishing $\tau^{-1}$. Both of the mentioned features are not observed in our numerical analysis.   
In Fig. 3, we distinguish the regions of validity (invalidity) of MOL Eq.(\ref{11}) by depicting the corresponding parts of
the curve representing this equation by dashed (dotted) segments. 
As can be elucidated from Fig.~3a), MOL theory does not cover the case $\tilde{\epsilon} < q^{-1}v_0$ corresponding to the $M-H$ boundary,
as this boundary is located to the left of the maximum of $v_m(\tau^{-1})$.  As is explained above,
our result for $v$ given by Eq.(\ref{10}) is free from the mentioned limitations as long as $k=q$ corresponds to the only marginally stable mode
of the spatial variation of the density. This fact explains the demonstrated validity of our results in a wide range of $\tau^{-1}$.

\section{Conclusions}
We have investigated the collective behavior of active SPPs described by their density and velocity orientation fields.
The three main effects are shown to influence the collective behavior of the SPPs, as follows:
(i) pattern-formation controlled by the parameters $\epsilon$ and $\bar{\rho}$,  being responsible for 
the significance of
pattern-forming effect and symmetry of the pattern morphologies, respectively;
(ii) orientational relaxation and dispersion of the SPP velocities controlled by the parameters $\tau$ and $\gamma$;
(iii) coupling between the density and orientational fields controlled by the parameter $v_0$.
As our main result we have shown, both analytically and numerically, that the form of  steady states of the SPP system
primarily depends on the significance of the orientational relaxation (orientational diffusion and dispersion) of the SPP velocity field.
In the absence of quick orientational relaxation of the SPP velocities, the individual motion of the SPPs results in a motion of the system 
in the direction of the shift of the polarization field relative to the density field of the SPPs.
With increasing significance of the orientational relaxation, the SPPs are forced to transit to either disordered or resting
ordered steady states, as is illustrated in Fig.2. Above a certain threshold magnitude of the orientational relaxation rate $\tau^{-1}(1+\gamma)$,
a spatial "synchronization" of the orientational field with the field of the SPP density gradients occurs. This synchronization
results in averaging out the velocities of the elementary SPP domains, thus suppressing their collective
motion.  If the significance of the pattern-forming effect
quantified through the parameter $\epsilon$ is not sufficient  to preserve the spatial structure of the patters in the whole range 
of the relaxation rates, the transition to the homogeneous steady state occurs at $\tau^{-1}\sim \tilde{\epsilon}$. In the frameworks
of the developed model, this case can be described by the inequality $\tilde{\epsilon} < q^{-1}v_0$. In the opposite case
$\tilde{\epsilon} > q^{-1}v_0$ of larger tendency to form patterns, the moving patterns experience the transition to the resting  state
without changing their density structure. This transition emerges through a monotonic decrease of the pattern velocity that drops to zero at the
transition boundary. Recall that the transition to the homogeneous state, in contrast, occurs at finite collective velocity of the SPPs. The amplitude
of the patterns, quantified through the RMSD defined above, behaves just  opposite to the collective velocities. Specifically, 
the amplitude tends to zero at the moving state-homogeneous state boundary, while having finite values at the moving state-resting state 
boundary. This interesting feature highlighting different behavior of the collective velocity and amplitudes of the patterns 
in the vicinity of the
transitions between different states, calls for a detailed non-linear analysis to reveal possible bifurcations 
between these states of the SPP system. Such an analysis will be reported elsewhere.

\end{document}